\documentclass[twocolumn,twocolappendix]{aastex63}

\usepackage{newtxtext,newtxmath}

\usepackage[T1]{fontenc}
\usepackage{ae,aecompl}
\usepackage{comment}
\usepackage{multirow}

\usepackage{graphicx,tabularx}	% Including figure files
\usepackage{amsmath}	% Advanced maths commands
\usepackage{amssymb}	% Extra maths symbols
\usepackage{booktabs}
\usepackage[inline,shortlabels]{enumitem}
\usepackage{bm}
\setlist{itemjoin* = { and\enspace}}

\newcommand{\be}{\begin{equation}}
\newcommand{\ee}{\end{equation}}
\newcommand{\bea}{\begin{eqnarray}}
\newcommand{\eea}{\end{eqnarray}}

\def\d{{\rm d}}

\newcommand{\refeq}[1]{Eq.~(\ref{eq:#1})}

\newcommand{\reffig}[1]{Fig.~\ref{fig:#1}}

\received{27-Apr-2020}
\revised{08-Nov-2020}
\accepted{16-Nov-2020}
\submitjournal{ApJ}

\newcommand{\HI}{H{\small I}}

\shorttitle{H{\small I} Bias Linearity}
\shortauthors{Z.~Wang et al.}
\graphicspath{{./}}
\begin{document}

% TITLE, AUTHORS AND STUFF -------------------------------------------------------------------
%\title{Can H{\small I} Gas Trace the Matter Density Distribution Linearly on Large Scales?}
\title{The Breakdown Scale of H{\small I} Bias Linearity}

\correspondingauthor{Yi Mao} \email{ymao@tsinghua.edu.cn}

\author[0000-0002-2970-3661]{Zhenyuan Wang}
\affiliation{Department of Astronomy, Tsinghua University, Beijing 100084, China}
\affiliation{Department of Astronomy and Astrophysics, The Pennsylvania State University, University Park, PA 16802, USA}

\author[0000-0002-4597-5798]{Yangyao Chen}
\affiliation{Department of Astronomy, Tsinghua University, Beijing 100084, China}
\affiliation{Department of Astronomy, University of Massachusetts, Amherst MA 01003-9305, USA} 

\author[0000-0002-1301-3893]{Yi Mao}
\affiliation{Department of Astronomy, Tsinghua University, Beijing 100084, China}

\author[0000-0001-5356-2419]{Houjun Mo}
\affiliation{Department of Astronomy, University of Massachusetts, Amherst MA 01003-9305, USA} 
\affiliation{Department of Astronomy, Tsinghua University, Beijing 100084, China}

\author[0000-0002-4911-6990]{Huiyuan Wang}
\affiliation{Key Laboratory for Research in Galaxies and Cosmology, Department of Astronomy, University of Science and Technology of China, Hefei, Anhui 230026, China}

\author[0000-0003-4936-8247]{Hong Guo}
\affiliation{Key Laboratory for Research in Galaxies and Cosmology, Shanghai Astronomical Observatory, Shanghai 200030, China}

\author[0000-0002-8711-8970]{Cheng Li}
\affiliation{Department of Astronomy, Tsinghua University, Beijing 100084, China}

\author{Jian Fu}
\affiliation{Key Laboratory for Research in Galaxies and Cosmology, Shanghai Astronomical Observatory, Shanghai 200030, China}

\author[0000-0002-4534-3125]{Yipeng Jing}
\affiliation{Department of Astronomy, and Tsung-Dao Lee Institute, Shanghai Jiao Tong University, Shanghai 200240, China}

\author[0000-0002-6593-8820]{Jing Wang}
\affiliation{Kavli Institute for Astronomy and Astrophysics, Peking University, Beijing 100871, China}

\author[0000-0003-3997-4606]{Xiaohu Yang}
\affiliation{Department of Astronomy, and Tsung-Dao Lee Institute, Shanghai Jiao Tong University, Shanghai 200240, China}

\author[0000-0003-1887-6732]{Zheng Zheng}
\affiliation{Department of Physics and Astronomy, University of Utah, 115 South 1400 East, Salt Lake City, UT 84112, USA}

% ABSTRACT AND KEYWORDS -----------------------------------------
\begin{abstract}
%The 21~cm intensity mapping experiments promise to measure the baryon acoustic oscillations (BAO) through the mapping of \HI\ gas at the post-reionization epoch. Its success depends on the assumption that \HI\ gas traces the matter density distribution {\it linearly} on large scales. However, both nonlinear halo clustering and nonlinear effects modulating \HI\ gas in halos may spoil this. 
The 21~cm intensity mapping experiments promise to obtain the large-scale distribution of \HI\ gas at the post-reionization epoch. In order to reveal the underlying matter density fluctuations from the \HI\ mapping, it is important to understand how \HI\ gas traces the matter density distribution. Both nonlinear halo clustering and nonlinear effects modulating \HI\ gas in halos may determine the scale below which the \HI\ bias deviates from linearity.
We employ three approaches to generate the mock \HI\ density from a large-scale N-body simulation at low redshifts, and demonstrate that the 
assumption of \HI\ linearity is valid at the scale corresponding to the first peak of baryon acoustic oscillations, but breaks down at $k \gtrsim 0.1\,h\, {\rm Mpc}^{-1}$. The nonlinear effects of halo clustering and \HI\ content modulation  
counteract each other at small scales, and their competition results in a 
model-dependent ``sweet-spot'' redshift near $z$=1 where the \HI\ bias is 
scale-independent down to small scales. We also find that the linear \HI\ bias 
scales approximately linearly with redshift for $z\le 3$.
\end{abstract}


%%%%%%%%%%%%%%%%% BODY OF PAPER %%%%%%%%%%%%%%%%%%

\section{Introduction}
Neutral hydrogen (\HI) atoms, which are expected to be 
contained in halos at low redshifts ($0.5\lesssim z \lesssim 3$), 
produce 21~cm line radiation that can be observed \citep{2010Natur.466..463C}. 
The 21~cm intensity mapping experiments, e.g., 
Tianlai\footnote{\url{http://tianlai.bao.ac.cn}}\citep{chen2012tianlai}, 
CHIME\footnote{\url{https://chime-experiment.ca}}\citep{bandura2014canadian}, 
HIRAX\footnote{\url{https://hirax.ukzn.ac.za}}\citep{newburgh2016hirax}, 
BINGO\footnote{\url{http://www.bingotelescope.org}}\citep{battye2013ra}, 
and SKA\footnote{\url{https://www.skatelescope.org}}\citep{2015aska.confE..12P}, 
which will survey the \HI\ mass distribution in 
very large volumes, provide a promising way to constrain the expansion history and 
structure formation in the Universe, thereby unveiling the nature of dark energy. 

These 21~cm intensity mapping experiments, despite low angular resolutions, can be used 
to detect large-scale features in the cosmological density field \citep{2008PhRvL.100i1303C,2008PhRvL.100p1301L}. 
For this purpose, it is important to understand how accurately \HI\ gas traces the matter density fluctuations. In general, the power spectrum of the \HI\ gas distribution is related to that 
of the underlying matter through a bias relation, 
$P_{\rm HI}(k) = b_{\rm HI}^2\,P_{\rm m}(k)$, where $b_{\rm HI}$ is the bias 
factor. It is, therefore, necessary to understand the bias factor,  
$b_{\rm HI}$, in particular its scale dependence, in order to use 
$P_{\rm HI}(k)$ to infer the distribution of mass in the universe. 
Note that the measurement of the baryon acoustic oscillations (BAO) can be obtained by using a template of wiggles in the power spectrum, which is least sensitive to the nonlinear bias. But the nonlinear bias can affect the broadband shape of power spectrum which also contains a wealth of cosmological information. 
In particular, it is important to determine the breakdown scale below which the \HI\ bias deviates from linearity, which is the focus of this paper. 
At quasi-linear scales, large-scale structure perturbation theory (see \citealt{2018PhR...733....1D,2020JCAP...05..005D} and references therein), which incorporates the higher-order bias parameters, may be developed to model the nonlinear \HI\ clustering (e.g.~\citealt{2019JCAP...09..024M}).

After cosmic reionization, most \HI\ gas is expected to be in galaxies, thanks to their high 
density and low temperature, while the neutral fraction in the intergalactic medium 
is very low, about $10^{-5}$. Furthermore, fluctuations in the ionization field are not 
expected to affect the \HI\ power spectrum on large scales \citep{2009MNRAS.397.1926W}.  
Thus, the distribution of the \HI\ gas may be understood in terms of its relation 
with galaxies, or with dark matter halos in which galaxies reside \citep{cai2016mapping,cai2017mapping,cui2017large}. 
Gas and star-formation processes can, in principle, change the 
\HI\ gas distribution in dark matter halos, and potentially introduce nonlinear 
bias in the relationship between \HI\ gas and dark matter \citep{2020ApJ...894...92G}.  
In addition, it is well-known from N-body simulations that the distribution of dark matter halos traces the underlying matter distribution nonlinearly at small scales \citep{2009ApJ...691..569J, 2013MNRAS.433..209N}. 
These nonlinearities, albeit at small scales (i.e., the size of halos), might 
spoil the \HI\ linearity assumption even on large scales, because of mode coupling 
on different scales. 

Previous studies of \HI\ bias either employed oversimplified \HI-halo mass relation 
(similar to the fitting formula in \citealt{2011MNRAS.415.2580K}) applied 
to N-body simulations \citep{2010MNRAS.407..567B,2012MNRAS.421.3570G,2016MNRAS.460.4310S,2016MNRAS.458..781P,2017MNRAS.464.4008P,2017MNRAS.469.2323P,2018MNRAS.476...96S}, 
or modelled the \HI\ gas using hydrodynamic simulations, such as  
IllustrisTNG \citep{2018ApJ...866..135V}, Illustris and Osaka \citep{2019MNRAS.484.5389A}. However, the volumes of gas simulations,  
typically $\lesssim (100\,h^{-1}\,{\rm Mpc})^3$, are usually too small 
to be valid on BAO scales ($\sim 100\,h^{-1}{\rm Mpc}$). 

Given its importance, in this paper, we study the relationship between \HI\ gas and 
dark matter on large scales, using three -- empirically, numerically, and observationally 
oriented, respectively -- approaches to model \HI\ gas in halos of different masses, 
and using halos in a large N-body simulation to construct the \HI\ gas distribution on large 
scales. Our simulation volume, $(500\,h^{-1}{\rm Mpc})^3$, is sufficiently large so that the finite 
box effect on the power spectrum and bias is negligible on BAO scales \citep{2019MNRAS.489.1684K}. The use of different models for HI\ gas
in halos also allows us to draw generic conclusions that are independent 
of our ignorance about the details of galaxy formation in dark matter halos.

The rest of this paper is organized as follows. In Section~\ref{sec:method}, we describe the methodology of modelling the \HI\ gas. We show the results and discussions in Section~\ref{sec:results}, and make concluding remarks in Section~\ref{sec:summary}.

\section{Mocking the \HI\ gas distribution} 
\label{sec:method}
Our \HI\ mock data is constructed from the results of a large-scale, high-resolution N-body simulation, {\it ELUCID} \citep{wang2016elucid}, of the $\Lambda$CDM universe, performed with the L-Gadget code, a memory-optimized version of Gadget-2 \citep{2005MNRAS.364.1105S}, in a comoving volume of $500\,h^{-1}{\rm Mpc}$ on each side using $3072^3$ particles. We refer the readers to \cite{wang2016elucid} for details of this simulation. To find halos, we use the FoF algorithm with a linking length of 0.2 times the mean particle separation. The SUBFIND algorithm \citep{springel2001} is employed to resolve the sub-structures (i.e.\ subhalos) in each FoF halo and to build the merger trees. We adopt an empirical model \citep{lu2014empirical} to construct the star formation histories of galaxies in those halos with masses above $10^{10} h^{-1}\,M_{\odot}$ (about 30 N-body particles). To fully trace the star formation history, we develop a Monte Carlo method to append unresolved progenitors to the leaf-halos of the merger tree \citep{2019ApJ...872..180C}. The \HI\ gas is then assigned to halos with masses above $10^{10} h^{-1}\,M_{\odot}$ using a star formation model \citep{krumholz2008atomic, krumholz2009atomic, krumholz2009star, krumholz2013star} that provides 
the full information about the star formation history. Finally, the \HI\ gas is smoothed onto grids to compute the \HI\ power spectrum. The key 
ingredients of our method are detailed below. The background cosmology is consistent with that 
given by the WMAP five-year data \citep{dunkley2009five}: 
$\Omega_{m}= 0.258$, $\Omega_{\Lambda} = 0.742$,  $\Omega_{b} = 0.044$, $h = 0.72$, $n_s = 0.96$ and $\sigma_8 = 0.8$.

\subsection{Star formation history}  
For resolved halos with $M_{\rm h} \ge 10^{10} h^{-1}\,M_{\odot}$, we follow the empirical model for star formation rate (SFR) as described in \cite{lu2014empirical} (their ``Model III''). The SFR of a {\it central} 
galaxy is assumed to depend only on the mass of its host halo, $M_{\rm h}$, and redshift $z$, 
\begin{equation}
{\rm SFR}(M_{\rm h},z)=\varepsilon\,\frac{f_b\,M_{\rm h}}{\tau}\,(X+1)^{\alpha}\,\left( \frac{X+R}{X+1}\right)^{\beta}\,\left(\frac{X}{X+R}\right)^{\gamma}\,. 
\label{eqn:SFR}
\end{equation} 
Here $\varepsilon$ is the overall efficiency, $f_b$ = $\Omega_{b}/\Omega_{m}$ is the cosmic baryon fraction, $\tau$ = $[1/(10H_0)](1+z)^{-3/2}$ describes the dynamical timescale of halos at a redshift $z$, 
the variable $X\equiv M_{\rm h}/M_c$ where $M_c$ is a characteristic mass scale. Other variables are 
parametrized as $\alpha=\alpha_0(1+z)^{\alpha'}$, and $\gamma = \gamma_{\rm a}$ if $z<z_c$, 
or, otherwise, $\gamma = (\gamma_{\rm a}-\gamma_{\rm b})[(z+1)/(z_{c}+1)]^{\gamma'}+\gamma_{\rm b}$. The free parameters ($\varepsilon$, $R$, $M_c$, $\alpha_0$, $\alpha'$, $\beta$, $\gamma_{\rm a}$, $\gamma_{\rm b}$, $\gamma'$, $z_c$) can be found by fitting the observed galaxy stellar mass functions and a composite local cluster conditional galaxy luminosity function at the $z$-band, as shown in \cite{lu2014empirical} (their Table 3). 
For unresolved halos with $M_{\rm h} < 10^{10} h^{-1}\,M_{\odot}$, Monte Carlo trees are adopted to extend their assembly histories down to $10^{9} h^{-1}\,M_{\odot}$ \citep{2019ApJ...872..180C}.

This model \citep{lu2014empirical} assumes that, during galaxy mergers, the SFR is under exponential 
decay in {\it satellite} galaxies where the gas can be stripped. As such, the \HI\ gas is dominated 
by the contributions from central galaxies. While this may not be true for big halos \citep{2018ApJ...866..135V}, we neglect the \HI\ gas from satellite galaxies, for simplicity.

%(\citep{2018ApJ...866..135V} pointed out that a little \HI\ can live outside the halos and \HI\ mass in satellite galaxies always dominates in big halos. However, we found that this discrepancy did not change the qualitative results below.)

With empirical star formation and merger models, we can trace the mass growth of each central galaxy from its merger tree, and obtain its stellar mass $M_*$. For a given halo mass, the stellar mass 
$M_*$ may not be the same in different halos because of their different merger histories.

\subsection{Star formation model} \label{subsec:SFR}
To connect the {\it surface} density of SFR $\dot{\Sigma}_*$ and that of gas mass $\Sigma_g$, 
we follow the star formation model developed in \cite{krumholz2008atomic, krumholz2009atomic, krumholz2009star, krumholz2013star}, 
\begin{equation}
\dot{\Sigma}_* = f_{\rm H_2}\,\epsilon_{\rm ff}\,\frac{\Sigma_{\rm g}}{{\rm t_{\rm ff} }}\,,
\end{equation}  
where $\epsilon_{\rm ff}$ = 0.01, ${\rm t}_{\rm ff} = 31 [\Sigma_g /(M_{\odot}\,\rm pc^{-2})]^{-0.25}\, {\rm Myr}$. 
Assuming that the gas is cold and comprised of ${\rm H_2}$ and \HI, the ${\rm H_2}$ fraction is given by 
\begin{equation}
f_{\rm H_2} = \left\{ \begin{array}{ll}
1-\frac{3}{4}(\frac{s}{1+0.25s})\,,  &  \textrm{if $s\leqslant2$} \\
0\,,						& \textrm{otherwise} 
\end{array} \right.
\end{equation}
The variable $s=\rm{ln}(1+0.6\chi+0.01\chi^2)/(0.6\tau_c)$, where 
$\tau_c = 320\,c\,Z_o \Sigma_g/({\rm g\,cm^{-2}})$, and the clumping factor $c = 1.0$. To estimate the gas phase metallicity relative to the solar one, $Z_o$, we adopt the average metallicity-stellar mass relation from the FIRE simulation \citep{ma2015origin}, 
${\rm log}{Z_o} = 0.35[{\rm log}(M_*/M_{\odot})-10] + 0.93\exp{(-0.43z)}-0.74$. 
%Some observations also provide the constrains of this relation to high redshifts(e.g. \cite{2008A&A...488..463M}). However, the minimal galaxy mass in observations only reaches 10$^8$M$_{\odot}$, while the smallest galaxy in the simulation can go below 10$^5$M$_{\odot}$. We need to point out that the scattering in the observation is often very large. The gas phase metallicity in small galaxies is the most uncertain parameter in this work. But our result is insensitive to the metallicity since small galaxies contribute little to the power spectrum, as we show in the section 3.2.
The radiation field parameter $\chi$ is estimated \citep{krumholz2013star} as  
$\chi = 72\,G_0'/n_{\rm CNM}$, where $G_0' = \dot{\Sigma}_* /\dot{\Sigma}_{*,0}$, $\dot{\Sigma}_{*,0} = 2.5\times10^{-3}M_{\odot}\,\rm pc^{-2}\, Myr^{-1}$, and $n_{\rm CNM}$ is the density of cold neutral medium (CNM) in units of ${\rm cm^{-3}}$. 
In molecular-poor regions, the CNM density is $n_{\rm CNM,hydro} \approx \Sigma_{\rm g}/(M_{\odot}\,\rm pc^{-2} )$, 
while in molecular-rich regions, the CNM density is $n_{\rm CNM,2p} = 72G_0' / [(3.1/4.1)(1+Z_o^{0.365})]$.  In general, $n_{\rm CNM} = \rm max\{n_{\rm CNM,2p}, n_{\rm CNM,hydro} \}$.

\subsection{Disk size}
To connect the surface density and the total density, we assume that the gas surface density follows an exponential profile, 
$\Sigma_g(r)=\Sigma_0{\rm e}^{-r/R_g}$.
%Some works argue that this assumption may not be true for galaxies with stellar mass below about 10$^9 \rm M_{\odot}$ at z = 0\citep{kelvin2014galaxy} and may overestimate the \HI\ mass in these galaxies\citep{popping2015evolution}. Again, as we will see in section 3.1, the \HI\ in small galaxies/halos contributes little to the cosmic \HI\ value. According to \citep{kravtsov2013size}, the cold gas disk is 2.6 bigger than the stellar disk size. 
We assume the gas disk to stellar disk size ratio $R_g/R_* = 3.3$ which fits best with the gas mass fraction in local galaxies \citep{lu2015galaxy} (c.f.\ $R_g/R_* = 2.6$ in \citealt{kravtsov2013size}). 
The stellar disk size at $z\approx 0.1$ is estimated \citep{dutton2011evolution} as
$R_*(M_*)=R_0(M_*/M_0)^{0.18}\big[(1/2)+(1/2)(M_*/M_0)^{1.8}\big]^{(0.34/1.8)}$, 
where $R_0 = 10^{0.72}\,{\rm kpc}$, $M_0 = 10^{10.44}\,M_{\odot}$. 
The disk size evolves with redshift as $R_{*}(z,M_*)=R_{*}(M_*)\,[(1+z)/1.1]^{-0.44}$.
%$\Delta{\rm log}_{10}(R_{*})=0.018-0.44{\rm log}_{10}(1+z)$. 

%\begin{figure}
%    \includegraphics[width=\columnwidth]{1EmpiricalSFR.eps}
%    \caption{Top: the empirical SFR in halos, based on the ELUCID simulation. Bottom: the central galaxy stellar mass vs halo mass. Shown are the mean (solid lines) at different redshifts, and the scattering at $z=0$ (gray dots).}    
%    \label{fig:EmpiricalSFR}
%\end{figure}
%%    %\emph{Upper panel:} the empirical SFR in dark matter halo predicted by \citep{lu2014empirical}. \emph{Lower panel:} central galaxy stellar mass(CGSM)-halo mass relation. The solid lines are the average CGSM-halo mass relation at different redshifts. The gray dots are the CGSM at $z$=0. The star formation rate is based on the modified halo merger trees in ELUCID simulation. 

\begin{figure}
    \includegraphics[width=\columnwidth]{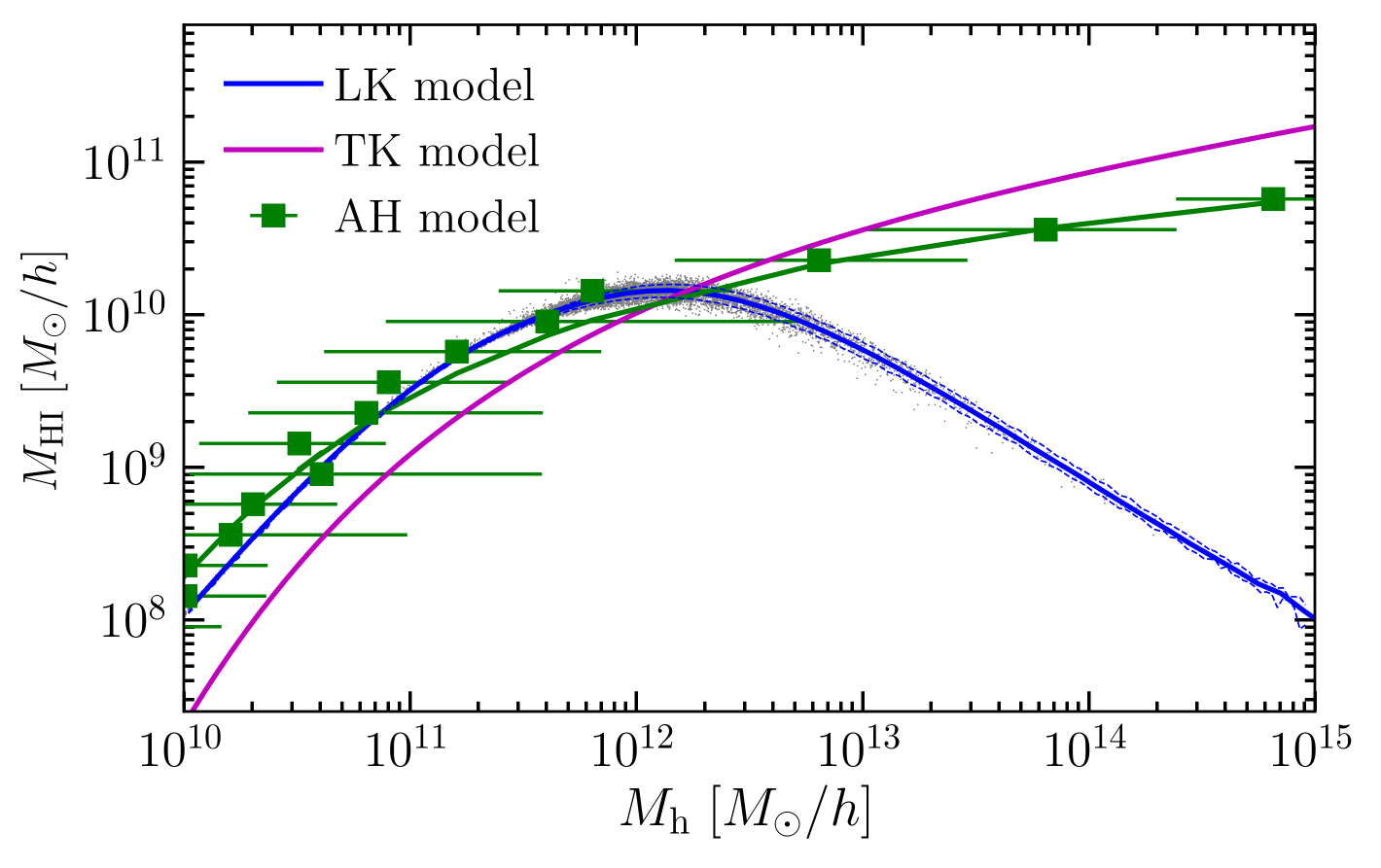}
    \caption{The \HI-halo mass relation derived from different models at $z=0$. We show the results using the LK model (blue), the TK model (magenta), and the AH model (green). Here we also include the scatter points (gray dots) and 1$\sigma$ envelope (blue dashed lines) for the LK model, and the error bars for the AH model.
    %Bottom: the fraction of cosmic \HI\ abundance in different halo bins with 0.1 dex bin width.
    } 
    \label{fig:HI-Halo}
\end{figure}

\subsection{\HI-halo mass relation}
In our above modelling, for a fixed stellar mass $M_*$, a given value of disk central density $\Sigma_0$ determines $\Sigma_g(r)$ at some radius in the disk. The aforementioned star formation model is employed to solve for $\dot{\Sigma}_*(r)$ numerically from $\Sigma_g(r)$, which gives the \HI\ surface density $\Sigma_{\rm \HI}(r)$. By integrating over the disk, we can find a correlation between the SFR and the \HI\ mass for a central galaxy, given $M_*$. For each halo, we compute the SFR using the aforementioned empirical model, and $M_*$ from halo merger history. Finally, the \HI\ mass is computed by interpolation using its correlation with SFR. Our \HI\ gas model, which incorporates the empirical SFR model \citep{lu2014empirical} and the star formation model \citep{krumholz2008atomic, krumholz2009atomic, krumholz2009star, krumholz2013star}, is dubbed ``LK model'', which stands for ``{\bf L}u et al.\ + {\bf K}rumholz et al.\ model''.

To test the model dependence of \HI\ bias, we also assign the \HI\ mass inside a halo by using the {\it average} \HI-halo mass relation obtained from two other approaches. One approach uses the IllustrisTNG simulation (their gas data)\citep{2018ApJ...866..135V} and the same star formation model\citep{krumholz2008atomic, krumholz2009atomic, krumholz2009star, krumholz2013star}. This model is dubbed ``TK model'' herein, which stands for ``Illustris{\bf T}NG + {\bf K}rumholz et al.\ model''. The average \HI-halo mass relation in the other approach was obtained by using the updated measurements of ALFALFA survey and HOD model \citep{2017ApJ...846...61G} (only available at $z=0$), and this model is dubbed ``AH model'', which stands for ``{\bf A}LFALFA data + {\bf H}OD model''. 
Following the fitting formula of average \HI-halo mass relation in \cite{2018ApJ...866..135V}, we use the following expression for both TK and AH models,
\begin{equation}
    \label{eq:fittingformula}
    M_{\mathrm{HI}}(M_{\rm h}, z)=M_{0}\left(\frac{M_{\rm h}}{M_{\min }}\right)^{\alpha} \exp \left[-\left(\frac{M_{\min }}{M_{\rm h}} \right)^{0.35}
    \right]\,.
\end{equation}
The bestfit parameter values, as listed in Table~\ref{tab:param}, are taken from \cite{2018ApJ...866..135V} (their Table 1 for FoF halos) for the TK model, and obtained by $\chi^2$-fitting the $M_{\rm HI}$-$M_h$ data at $z=0$ for the AH model.

\begin{table}
    \centering
    \caption{Parameter values used for the TK and AH models.}
    \begin{tabular*}{7cm}{ccccc}
    \hline\hline
    Model & $z$ & $\alpha$ & $M_{0}[M_{\odot}/h]$ & $M_{\min } [M_{\odot}/h]$   \\ 
    \hline
    \multirow{4}{*}{TK} &0 & 0.24 & $4.3 \times 10^{10}$ & $2.0 \times 10^{12}$ \\
    {} & 1 & 0.53 & $1.5 \times 10^{10}$ & $6.0 \times 10^{11}$ \\
    {} & 2 & 0.60 & $1.3 \times 10^{10}$ & $3.6 \times 10^{11}$\\
    {} & 3 & 0.76 & $2.9 \times 10^{9}$ & $6.7 \times 10^{10}$ \\
    \hline 
    \multirow{1}{*}{AH} & 0 & 0.12 & $2.6 \times 10^{10}$ & $6.9 \times 10^{11}$ \\
    \hline
    \end{tabular*}
    \label{tab:param}
\end{table}

In Figure~\ref{fig:HI-Halo}, we show the \HI-halo mass relation for central galaxies at $z=0$. Our results (LK model) are compared with predictions from the IllustrisTNG simulation (TK model), and the results from updated ALFALFA observations (AH model). All results agree well for low-mass halos ($M_{\rm h} < 10^{11}h^{-1}\,M_{\odot}$). We checked that this agreement holds well at higher redshifts ($0<z<2$) between LK and TK models. For massive halos, nevertheless, our model underestimates the \HI\ mass, for two possible reasons. First, the \HI\ mass in the TK model includes the contributions from both central and satellite galaxies, while both our model and the AH model only consider those from the central galaxies. Secondly, our empirical model might underestimate the SFR for massive halos. However, the contribution of \HI\ gas from massive halos is generally not important due to the sharp decrease of the halo mass function towards the massive end. In addition, the slope of \HI-halo mass curve declines at the high mass end, which further suppresses the contribution of \HI\ gas inside the massive halos. We will further discuss the impact of \HI\ modelling in the high-mass end on the linear \HI\ bias in Section~\ref{subsec:linearHI} below.

\subsection{\HI\ Power spectrum} 
The \HI\ mass in each halo is smoothed onto a uniform grid with $1024^3$ cells, and we compute the \HI\ power spectrum from the FFT. We only keep the power spectrum for wavenumber less than a quarter of Nyquist number ($k < 1.57h\,{\rm Mpc}^{-1}$) to avoid the alias effect. 
%can be expanded in polynomials of the local matter overdensity field on large scales , 
%\begin{equation}
%\delta_{\rm HI}({\bf x}) = b_1\,\delta_{\rm m}({\bf x}) +  \frac{1}{2}\,b_2\,\delta_{\rm m}^2({\bf x}) + \epsilon({\bf x}) + \ldots
%\end{equation}
In Fourier space, we can define a scale-dependent effective bias, $b_{\rm HI}(k)$,
\begin{equation}
\delta_{\rm HI}(\bm{k}) = b_{\rm HI}(k)\,\delta_{\rm m}(\bm{k}) + \epsilon(\bm{k})\,,
\end{equation} 
where $\epsilon({\bm k})$ is a stochastic component which does not 
correlate with the density field, $\delta_{\rm m}$. 
On large scales, we expect $b_{\rm HI}$ is a scale-independent linear bias. 

The \HI\ bias can be estimated using the auto-power spectrum of \HI\ gas, 
$b^{\rm uncorr}_{\rm HI,auto}(k) = [P_{\rm HI}(k)/P_{\rm{m}}(k)]^{1/2}$, 
if the shot noise is uncorrected. 
The leading-order mass-weighted \HI\ shot noise is estimated by shuffling \HI\ gas randomly, i.e.\ $P_{\rm SN} = V_{\rm survey}^{-1}\langle \epsilon({\bm k})\epsilon(-{\bm k}) \rangle$, and then subtracted from the raw power spectrum. After correcting for shot noise, we have 
\begin{equation}
\label{eqn:bias_auto}
b_{\rm HI,auto}(k) = \sqrt{\frac{P_{\rm HI}(k) - P_{\rm SN}}{P_{\rm m}(k)} } \,. 
\end{equation}
The assumption of \HI\ linearity can be tested by checking if the \HI\ bias, $b_{\rm HI,auto}(k)$, is equal to the scale-independent linear bias at large scales. 
Of course, the \HI\ bias is expected to be scale-dependent at small scales 
due to nonlinear evolution. 

  The \HI\ bias may also be estimated using the cross-power spectrum between \HI\ density and total matter density, 
\begin{equation}
    b_{\rm HI,cross}(k) = \frac{P_{\rm HI,m}(k)}{P_{\rm m}(k)}.
    \label{eqn:bias_cross}
\end{equation}
This estimator avoids the shot noise automatically. However,  in this paper, we choose to estimate the \HI\ bias based on the auto-power spectrum of \HI\ gas, because the 21~cm intensity mapping measures the auto-power spectrum of the 21~cm brightness temperature. 
As shown below in Section \ref{subsec:def-bias}, the results from these two estimators 
are in good agreement. Thus we neglect the subscript ``auto'' throughout this paper 
except in  Section \ref{subsec:def-bias}.

\begin{figure*}
    \centering
    \includegraphics[width=2\columnwidth]{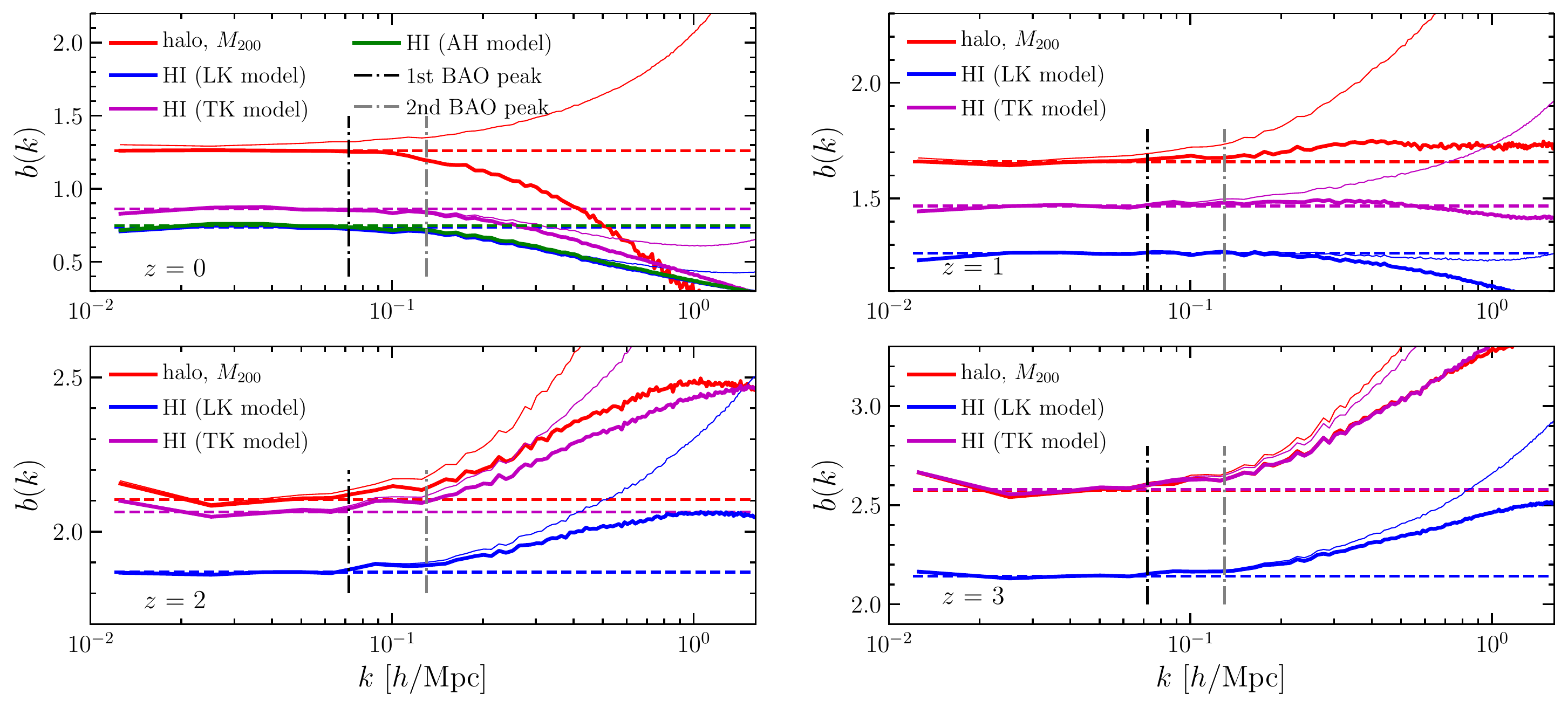}
    \caption{The bias of halo mass density fluctuations (red) and \HI\ mass density fluctuations derived from the LK(blue), TK(magenta), and AH(green) models 
    at $z=0$, 1, 2, and 3, respectively, with respect to the matter density fluctuations, with shot-noised corrected (thick solid lines) and uncorrected (thin solid lines). The dashed lines indicate the constant linear bias which is estimated by averaging over $k = 0.025$ --- 0.075 $h\,{\rm Mpc}^{-1}$ (we neglect the smallest $k$-mode due to its relatively large cosmic variance). The dot-dashed vertical lines mark the wavenumbers of the first (black) and second (grey) BAO peaks. }
    \label{fig:PS99}
\end{figure*}

\section{Results and Discussion}
\label{sec:results}
\subsection{Generic behavior}
In Figure~\ref{fig:PS99}, we show the \HI\ bias from different \HI-halo mass relations at different redshifts (except that the AH model is only at $z=0$) as well as the halo bias. 
In all three models, the \HI\ bias remains a constant at large scales for $k\lesssim 0.1\,h\,{\rm Mpc}^{-1}$, i.e.\ we confirm that, generically, \HI\ gas is indeed a linear biased tracer at the first BAO peak. However, the linearity assumption begins to break down at the second BAO peak. To test whether this break-down scale relies on the halo resolution in our simulation, we vary the minimum halo mass from $10^{10} h^{-1}\,M_{\odot}$ to $10^{11} h^{-1}\,M_{\odot}$, and find that while the amplitude of \HI\ bias depends on the halo mass cutoff similar to that of the halo bias, the linearity break-down scale is almost unchanged. 
Also, to test the effect of satellite galaxies, we estimate the \HI\ masses from satellite galaxies and assign them to the centers of subhalos, using the LK model at $z=0$. 
We find that including satellites does not change the shape of the \HI\ power spectrum significantly on scales $k\lesssim 1\,h\,{\rm Mpc}^{-1}$.

\begin{figure}
    \includegraphics[width=\columnwidth]{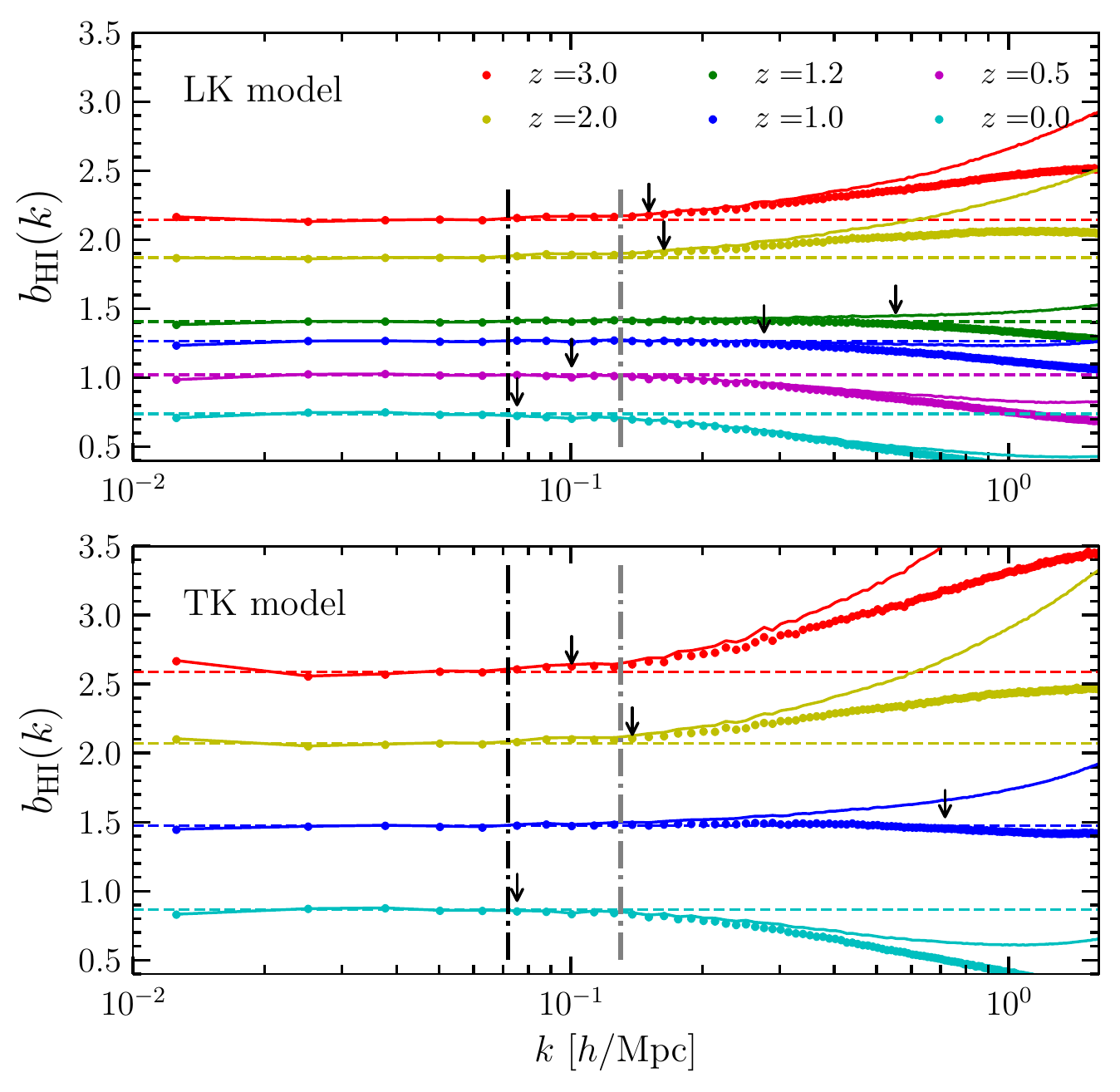}
    \caption{The redshift evolution of the \HI\ bias from $z=0$ to $3$, from the LK model (top) and the TK model (bottom), with shot-noised corrected (thick solid lines) and uncorrected (thin solid lines). The dashed lines indicate the constant linear bias 
    which is estimated by averaging over $k = 0.025$ --- 0.075 $h\,{\rm Mpc}^{-1}$. The arrow marks the scale at which the \HI\ bias deviates from the linear bias at the 1.5\% level. The dot-dashed vertical lines mark the scales of the first (black) and second (grey) BAO peaks.}
    \label{fig:scale bias}
\end{figure}

The behaviors at small scales are more interesting,
as most of the \HI\ gas resides only inside halos after cosmic reionization. Figure~\ref{fig:PS99} shows that nonlinear halo clustering always enhances the halo power spectrum at small scales (before corrected for shot noise). However, Figure~\ref{fig:HI-Halo} shows that \HI\ mass is suppressed in large halos. This suppression decreases the \HI\ density fluctuations at small scales relative to the level of fluctuations caused by halos
(see Fig.\,~\ref{fig:PS99}). The \HI\ suppression effect is stronger at lower 
redshifts as more massive halos form. The competition between these two opposite effects, 
namely the nonlinear effects in halo clustering and those modulating 
the \HI\ gas in halos, determines the evolution of the \HI\ bias at small 
scales. As shown in Figure~\ref{fig:scale bias}, 
for both LK and TK models, the \HI\ bias at small scales is enhanced with 
respect to the linear bias at high redshifts, just like the nonlinear halo bias, 
while the \HI\ bias is actually suppressed at small scales at $z=0$. 

The halo bias is known to become scale-dependent at $k\gtrsim 0.1\,h\,{\rm Mpc}^{-1}$ \citep{2009ApJ...691..569J, 2013MNRAS.433..209N} from N-body simulations. Naively, this sets the generic scale for the breakdown of linearity in \HI\ bias, since most of the 
\HI\ gas resides inside halos after cosmic reionization. Nevertheless, the nonlinearity of the \HI\ content significantly affects the level of \HI\ fluctuations with respect 
to halo biasing, thereby modulating the breakdown scale and making it
redshift-dependent, as shown in Figure~\ref{fig:scale bias}.
In particular, these two nonlinear effects appear to balance each other 
at a transition time where the \HI\ bias is linear down to small scales. 
In the LK model, this ``sweet-spot'' redshift is near $z=1.2$, with the linearity 
extending down to a scale $k\simeq 0.5\,h\,{\rm Mpc}^{-1}$. In the TK model, 
the transition takes place at $z\approx 1$, with the linearity extending 
down to $k\simeq 0.7\,h\,{\rm Mpc}^{-1}$. Thus, the ``sweet-spot'' redshift 
is likely to be near $z=1$, although the exact value is model-dependent.

%In Figure~\ref{fig:scale bias}, we find that, for both LK and TK models, while the \HI\ density at small scales is suppressed with respect to the linear bias near $z=0$, the \HI\ bias becomes enhanced at higher redshifts. The competition between two opposite effects results in the evolution of \HI\ bias at small scales. \HI\ gas can be held only inside halos after cosmic reionization, and halo mass density fluctuations are always enhanced at small scales (before corrected for shot noise). On the other hand, \HI\ mass is suppressed (i.e.\ $dM_{\rm HI}/dM_{\rm h}$ turns small) in large halos, which decreases the \HI\ density fluctuations at small scales with respect to the level of fluctuations caused by halos. The \HI\ suppression effect is stronger at low redshifts because more massive halos are formed. When these two effects are balanced out, there exists a transition time when the \HI\ bias is constant and linear down to some small scales. In the LK model, we find that this ``sweet-spot'' redshift is near $z=1.2$ with the linearity break-down at $k\simeq 0.5\,h\,{\rm Mpc}^{-1}$. In the TK model, the transition takes place at $z\approx 1$. As such, the ``sweet spot'' redshift may contain the information of the \HI-halo mass relation.

\begin{figure}
    \includegraphics[width=0.95\columnwidth]{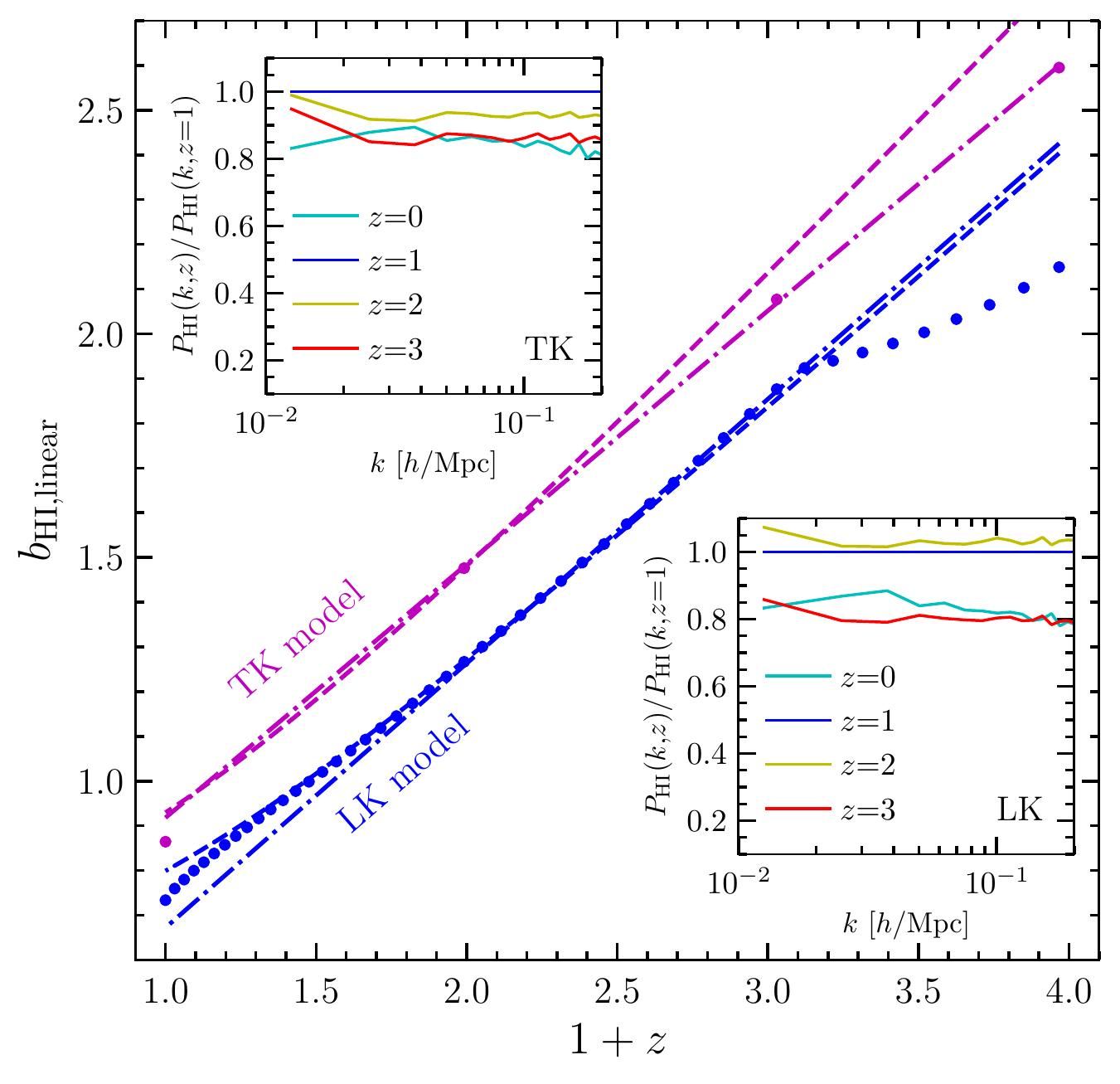}
    \caption{The redshift evolution of the \HI\ \emph{linear} bias in the LK model (blue dots) and the TK model (magenta dots). We fit the data linearly between $z=1$ and 2 (dot-dashed lines). For diagnostic purpose, we plot the prediction of linear \HI\ bias if \HI\ density power spectra at different redshifts would be the same as in $z=1$ but matter density fluctuations evolve according to linear theory (dashed lines). We also plot the ratio $P_{\rm HI}(k,z)/P_{\rm HI}(k,z=1)$ for both models in insets.}
    \label{fig:linear evolution}
\end{figure}

\begin{figure}
    \includegraphics[width=0.95\columnwidth]{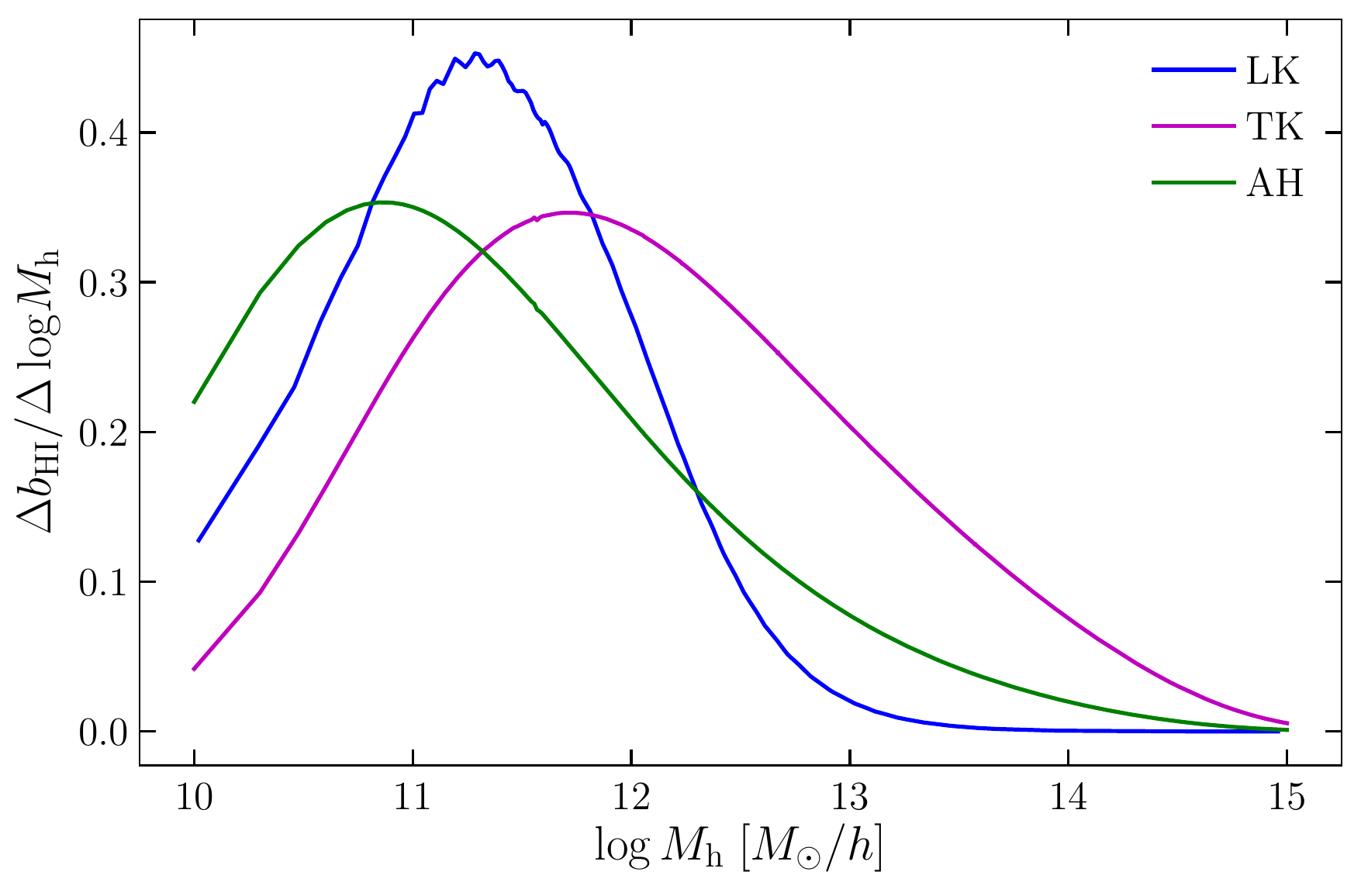}
    \caption{The contribution to the linear \HI\ bias from different logarithmic halo mass bin for the LK (blue), TK (magenta), and AH (green) model at $z = 0$. }
    \label{fig:BiasContribution}
\end{figure}

\subsection{Linear \HI\ bias}
\label{subsec:linearHI}
The linear \HI\ bias (i.e.\ the constant \HI\ bias averaged over large scales) increases with redshift, as shown in Figure~\ref{fig:linear evolution}. We find an interesting feature in 
both LK and TK models. In general, the \HI\ bias varies \emph{approximately} 
linearly with redshift. This linear relation is almost exact between $z = 1$ and 2, 
with an error $<10\%$ for $z<1$ and $<15\%$ for $2<z<3$. This can be understood as follows.
The linear \HI\ bias can be written as $b_{\rm HI,linear}(z) =  \big[D_{\rm HI}(z)/D_{\rm m}(z)\big]\, b_{\rm HI,linear}(1)$, where $D_{\rm HI}$ and $D_{\rm m}$ are the linear growth functions of the \HI\ and matter density fluctuations, respectively, i.e.\ $D_{\rm HI}(z)=\big[P_{\rm HI}(z)/P_{\rm HI}(1)\big]^{1/2}$ and $D_{\rm m}(z)=\big[P_{\rm m}(z)/P_{\rm m}(1)\big]^{1/2}$.
As shown in the insets of Figure~\ref{fig:linear evolution}, the \HI\ density power spectrum varies only slightly with redshift, i.e.\ $D_{\rm HI}(z)\approx 1$. 
The similar result was also found in \cite{2018ApJ...866..135V}. The reason that \HI\ clustering only weakly varies at $0<z<3$ is an interesting open question.
On the other hand, in a matter-dominated universe, the matter growth function scales as $D_{\rm m}(z)\propto (1+z)^{-1}$ \citep{cooray2002halo}. 
These two effects combined lead to the linear scaling relation, 
$b_{\rm HI,linear}(z) \propto (1+z)$, which we find to be generic.\footnote{Coincidentally, the linear galaxy bias also typically scales linearly with $1+z$, because for a passively evolving population, $b_{\mathrm{gal}}(z)-1= [b\left(z_{0}\right)-1] D\left(z_{0}\right)/D(z)$ \citep{Fry_1996,Skibba_2014}, and in a matter-dominated universe, $D(z)\propto (1+z)^{-1}$. However, this cannot explain the nearly linear scaling of \HI\ bias evolution we find herein, because the above relation only holds for a tracer with conservative total number, i.e.\ a passively evolving population, and therefore the bias is predicted to be either always greater or always smaller than unity. But Figure~\ref{fig:linear evolution} shows that the linear \HI\ bias crosses the unity between $z=0$ and $z=1$ for both LK and TK model.}
There are two reasons why this relation is not exactly linear. 
First, $D_{\rm m}(z)$ is suppressed at $z<1$ when dark energy kicks in. 
Secondly, the \HI\ power spectrum has small, non-monotonous, evolution with redshift. 
As an illustration, consider a case in which $D_{\rm HI}(z)=1$, but 
$D_{\rm m}(z)$ takes the value from the linear perturbation theory 
(including the effect of dark energy). We find that the prediction of 
the linear \HI\ bias in this case agrees with the actual results in both models, 
%More precisely, the former can be slightly higher than the latter 
with $< 15\%$ error. 
This is consistent with the fact that the \HI\ power spectrum reaches its 
maximum at $z\simeq 1-2$, with the values at $z=0$ and $3$ about 20\% lower 
than the maximum.

Other than the generic results presented above, however, the value of the 
linear \HI\ bias can be model-dependent. Figures~\ref{fig:PS99}--\ref{fig:linear evolution} show that in general the TK model predicts a higher value of linear \HI\ bias 
than the LK and AH models. This difference might be attributed to the contributions 
of the \HI\ gas in massive halos. 
We can understand this with halo model, in which the linear bias can be written as the integration of contributions from halos with different mass, 
\bea
b_{\mathrm{HI,linear}}(z)=\frac{\int_{M_{\rm min}}^{M_{\rm max}} n(M_{\rm h}, z)\, b(M_{\rm h}, z)\, M_{\mathrm{HI}}(M_{\rm h}, z) \,d M_{\rm h}}{\int_{M_{\rm min}}^{M_{\rm max}} n(M_{\rm h}, z)\, M_{\mathrm{HI}}(M_{\rm h}, z) \,d M_{\rm h}}.\nonumber
\eea
We calculate the prediction of linear \HI\ bias in halo model using the fitting formula of the halo bias $b(M_{\rm h},z)$ and halo abundance $n(M_{\rm h}, z)$ in \cite{tinker2008toward,tinker2010large}, and the average \HI-halo mass relation for all three models at $z = 0$, and find the results agree quite well with the bias directly measured from the simulation. In \reffig{BiasContribution}, we show the contribution to the \HI\ bias from each logarithmic halo mass bin of finite stepsize,  
\bea
\frac{\Delta b_{\rm HI}(M_{\rm h})}{\Delta \log M_{\rm h}} = \frac{n(M_{\rm h}, z) b(M_{\rm h}, z) M_{\mathrm{HI}}(M_{\rm h}, z) \left(\Delta M_{\rm h}/\Delta \log M_{\rm h}\right)}{ \int_{M_{\rm min}}^{M_{\rm max}} n(M_{\rm h}, z) M_{\mathrm{HI}}(M_{\rm h}, z) \d M_{\rm h}}.
\nonumber
\eea
(For the $n^{\rm th}$-bin, $\Delta M = M_{n + 1} - M_n$, $\Delta \log M = \log M_{n + 1} - \log M_n$.)
For all three models, \reffig{BiasContribution} shows that the peak contribution appears at $M_{\rm h} = 10^{11} - 10^{12}\,M_{\odot}/h$, i.e.\ the intermediate-mass halos contribute most to the linear \HI\ bias. If we add up the contributions from different halo mass bins, we find that the massive halos of $M_{\rm h} = 10^{14} - 10^{15}\,M_{\odot}/h$ only contribute to $3.7\%$ of the linear \HI\ bias in TK model, but contribute to about $30\%$ of the linear halo bias. This indicates that the decreasing slope of the \HI-halo mass relation at the high mass end indeed further suppresses the linear \HI\ bias. Since the \HI\ mass is more suppressed in the massive halos in the LK and AH model than in the TK model, this explains why the linear \HI\ bias is smaller in the former. 
We also point out that since the contribution at our lower mass limit $M_{\rm h} = 10^{10}M_{\odot}/h$ does not vanish in \reffig{BiasContribution}, especially for the LK and AH model, our results of linear \HI\ bias may be overestimated due to the neglect of unresolved smaller-mass halos which smooth out the fluctuations.

%The lower panel of Figure~\ref{fig:HI-Halo} shows that the dominant contribution in all models is from halos with mass around $10^{11} h^{-1}\,M_{\odot}$. While the contributions from more massive halos in both LK and AH models drop rapidly, the corresponding contribution in the TK model is much higher. This is consistent with the fact that each massive halo in the TK model can generate more \HI\ gas than that in the other two models. Since massive halos, e.g.\ around $10^{13} h^{-1}\,M_{\odot}$, are rare, these massive halos cause a larger fluctuation in \HI\ density in the TK model.

\begin{figure}
    \includegraphics[width=\columnwidth]{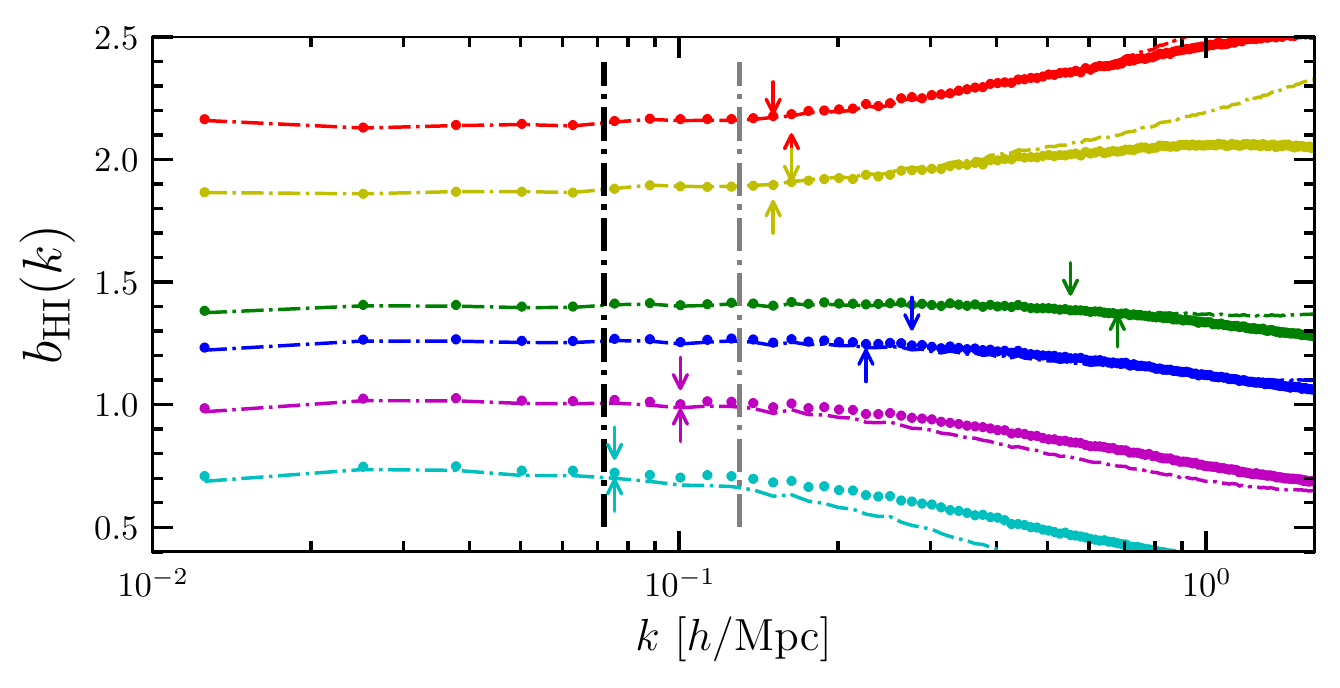}
    \caption{The \HI\ bias defined by auto-power spectrum after correcting for shot noise, $b_{\rm HI,auto}$ (dots) (see equation~\ref{eqn:bias_auto}), and cross-power spectrum, $b_{\rm HI,cross}$ (dot-dashed lines) (see equation~\ref{eqn:bias_cross}), at various redshifts $z=3.0$, 2.0, 1.2, 1.0, 0.5, and 0 (for dots and lines from top to bottom, respectively) in the LK model. The upward and downward arrows mark the scale at which the \HI\ bias $b_{\rm HI,cross}$ and $b_{\rm HI,auto}$ deviate from the linear bias at the 1.5\% level, respectively.}  
    \label{fig:bias_auto_cross}
\end{figure}

\subsection{Auto- vs. cross-power spectrum}
\label{subsec:def-bias}

In Figure~\ref{fig:bias_auto_cross}, we compare the \HI\ bias obtained from the auto-power 
spectrum with that obtained from the cross-power spectrum. The two results agree very well with each other down to very small scales.  
We note that the small difference between them does not affect 
any of the conclusions reached above. 

\subsection{Comparison with previous work}

\cite{2018MNRAS.473.4297P} performed an analytical calculation that accounts 
for the contribution from nonlinear matter fluctuations and nonlinear 
\HI\ modulation, using the combination of a perturbation theory 
and a halo model. They employed six different fitting formulae for the 
\HI-halo mass relation at $z=1$. Their results show that the \HI\ bias 
is scale-independent in the range $k=0.01 - 0.1\,h\,{\rm Mpc}^{-1}$. 
However, their results indicate that the \HI\ bias is weakly 
scale-dependent at $k\sim 0.001\,h\,{\rm Mpc}^{-1} $ and significantly 
scale-dependent at $k> 0.1\,h\,{\rm Mpc}^{-1}$ (see their figure 4), 
which is different from our results. Because of the limitation of
our simulation volume, our results are reliable only for 
$k>0.01 \,h\,{\rm Mpc}^{-1}$, which makes it difficult to test 
the presence of scale-dependence on ultra-large scales. The redshift $z=1$ in our 
results is near the sweet-spot redshift where the \HI\ bias is scale-independent 
down to scales smaller than $k\sim 0.1 \,h\,{\rm Mpc}^{-1}$. 
The difference on small scales may be due to the different 
methodologies adopted in the two investigations.
While perturbative calculations can provide important insights, 
numerical simulations can account for nonlinear effects more accurately. 

\cite{2016JCAP...03..061U} and \cite{2018MNRAS.473.4297P} 
investigated nonlinear effects in {\it observations}, such as nonlinear 
redshift-space distortion and nonlinear lensing, based on perturbative 
calculations, and found that these nonlinearities can also produce 
scale-dependent bias on large scales. We will explore these effects in 
numerical simulations in the future.

\cite{2020MNRAS.493.5434S}\footnote{While \cite{2020MNRAS.493.5434S} published earlier than us, our original preprint was posted on arXiv eight months earlier.} investigated the \HI\ content in halos using N-body simulations and semi-analytical model of galaxy evolution and gas content. This approach is similar to ours but the two 
differ in details. Their simulation box is the same as ours, 
both in a comoving volume of $500\,h^{-1}\,{\rm Mpc}$ on each side. 
They found that the \HI\ bias is scale-independent on large scales,
which is similar to ours, but their results are much noisier (see their Figure 12). 
They also found that the \HI\ bias is enhanced at high redshifts. 
However, the \HI\ bias in their results is roughly scale-independent down 
to $k\sim 2\,h\,{\rm Mpc}^{-1}$ up to $z=2$. In contrast, our results show 
that the small-scale bias evolves from being enhanced at higher redshift
to being suppressed at $z<1$, and that there is a sweet-spot redshift 
near $z=1$. This difference is likely due to different treatments in
the \HI\ contents of halos, which can lead to different suppression 
of the nonlinear \HI\ modulation.

\cite{2018ApJ...866..135V} made a comprehensive analysis of \HI\ gas distribution based on the IllustrisTNG megneto-hydrodynamic simulation. They employed the similar gas model \citep{krumholz2008atomic, krumholz2009atomic, krumholz2009star, krumholz2013star} to ours\footnote{Nevertheless, since the atomic hydrogen is the dominant component in the cold gas, the \HI\ content inside halos is mostly determined by the gas physics in hydro simulations, and less affected by the gas model during the post-processing. To see this, \citealt{diemer2018modeling} applied different cold gas model to the same hydro simulation, but found the similar \HI\ gas mass in galaxies in their Figure 4.}, 
to divide the cold hydrogen gas within each cell into atomic and molecular components.  
Due to the limited simulation volume ($\sim 100\,h^{-1}\,{\rm Mpc}$ on each side), \cite{2018ApJ...866..135V} cannot test the scale-dependence of the \HI\ bias at the first BAO peak scale. 
In comparison, our work applied the average \HI-halo mass relation from \cite{2018ApJ...866..135V}, nevertheless, to halos resolved from our N-body simulation with large enough volume ($500\,h^{-1}\,{\rm Mpc}$ on each side) --- i.e.\ our TK model --- to avoid the finite box effect on the bias at the BAO scales. Therefore, we can directly confirm from simulation that the \HI\ bias is linear at the scale corresponding to the first BAO peak. On the other hand, both \cite{2018ApJ...866..135V} and our work find that the \HI\ bias linearity becomes to break down at $k \gtrsim 0.1 h\,{\rm Mpc}^{-1}$ generically, the smallest wavenumber presented in \cite{2018ApJ...866..135V}.  

\cite{2019JCAP...09..024M} investigated the clustering of \HI\ gas using the {\it Hidden Valley} N-body simulation which has a large comoving volume of 1 $h^{-1}\,{\rm Gpc}$ on each side at $z = 2-6$, so their work is complementary to ours regarding the focused regime of redshift. The Hidden Valley simulation can resolve halos down to $10^9 M_\odot/h$, so they can incorporate the \HI\ gas inside smaller-mass halos at high redshifts than our work. They adopted fitting formulae of average \HI-halo mass relation similar to \refeq{fittingformula} herein, in order to assign \HI\ mass to halos, and explore both two-point correlation function and power spectrum. Similar to our results, they also found that the \HI\ bias becomes scale-dependent at $k \gtrsim 0.1 h\,{\rm Mpc}^{-1}$. Their results indicate a strong scale-dependence of \HI\ bias at the large $k$ at the high redshift, which is consistent with the behavior in our results for the redshifts higher than the sweet-spot $z\approx 1$.

\section{Summary}
\label{sec:summary}
In this paper, we use a large N-body simulation to explore the \HI\ bias 
for 21~cm intensity mapping experiments at low redshifts. We adopt three models, LK, TK and AH, 
representing empirically, numerically, and observationally 
oriented approaches, respectively, to assign \HI\ mass to  
dark matter halos and to account for uncertainties in the 
\HI-halo mass relation. 

We confirm that the \HI\ gas distribution is a linearly biased tracer 
of the total dark matter density field on the scales corresponding to the first BAO peak. 
However, the \HI\ linearity assumption breaks down at $k > 0.1\,h\,{\rm Mpc}^{-1}$. The exact breakdown scale is redshift-dependent, because the nonlinear effects that modulate the \HI\ gas in halos evolve with time. This \HI\ nonlinearity, which 
is caused by the nonlinear halo clustering and nonlinear \HI\ content modulation, 
is {\it intrinsic} and not related to the instrumental and observational effects. 
This imposes a challenge to the upcoming 21~cm intensity mapping experiments 
in their capabilities to extract cosmological information from 
the broadband shape of the 21~cm power 
spectrum in this $k$-range where a large number of modes are located. 
The result is particularly important for forecasting  
cosmological constraints with upcoming 21~cm intensity 
mapping experiments. 
It is, therefore, necessary to better model the \HI\ power spectrum beyond the linear regime, e.g. applying the large-scale structure perturbation theory at the quasi-linear scales. 
We note, however, that cosmological constraints from the BAO measurement of the 21~cm power spectrum is not affected by the nonlinear bias.

We find the existence of a characteristic redshift above and below 
which the small scale \HI\ bias is enhanced and suppressed
relative to the linear bias, respectively. 
For redshifts close to this ``sweet spot'', the \HI\ bias is linear 
down to small scales. For example, for the LK model, the characteristic 
redshift is $\simeq 1.2$, at which the linearity of the bias 
extends from large scales all the way down to
$k\simeq 0.5\,h\,{\rm Mpc}^{-1}$.  
However, the exact value of this ``sweet spot" redshift depends 
both on the \HI-halo mass relation and on nonlinear clustering of halos. 
Determining the ``sweet-spot" redshift observationally can, therefore, 
also provide valuable information on star formation and clustering of dark matter halos.  
%But this diagnostic is likely to be less useful, in practice, since there are too many uncertain parameters altogether. 
%This may have twofold impacts. On one hand, a much larger number of modes down to small scales can be taken to put more stringent constraints on cosmological parameters. On the other hand, the exact redshift of this epoch might be sensitive to the \HI-halo mass relation. Therefore, pinning down the sweet spot redshift observationally may place constraints on the star formation model. 

Finally, we also find that the linear \HI\ bias is an approximately linear function of 
redshift for $z\le 3$. This may make cross-checks between different 
redshifts more powerful for interpreting observational data.

\section*{Acknowledgements}
This work is supported by the National Key R\&D Program of China (Grant No.2018YFA0404502, 2018YFA0404503, 2017YFB0203302),
and the National Natural Science Foundation of China (NSFC Grant No.11673014, 11761141012, 11821303, 11543006, 11833005, 11828302, 11922305, 11733004, 11773049, 11761131004, 11673015, 11421303, 11721303, U1531123).  
YM and JW were also supported in part by the Chinese National Thousand Youth Talents Program. 
JF acknowledges the support by the Youth innovation Promotion Association CAS and Shanghai Committee of Science and Technology (Grant No.19ZR1466700).
HJM was also supported in part by the NSF (Grant No. AST-1517528). 
We are grateful to Xuelei Chen, Kai Hoffmann, Adam Lidz, Matt McQuinn and Francisco Villaescusa-Navarro for useful discussions, and the anonymous referee for constructive comments.

\bibliographystyle{aasjournal}
\bibliography{My_Collection} {}

\begin{thebibliography}{}
\expandafter\ifx\csname natexlab\endcsname\relax\def\natexlab#1{#1}\fi
\providecommand{\url}[1]{\href{#1}{#1}}
\providecommand{\dodoi}[1]{doi:~\href{http://doi.org/#1}{\nolinkurl{#1}}}
\providecommand{\doeprint}[1]{\href{http://ascl.net/#1}{\nolinkurl{http://ascl.net/#1}}}
\providecommand{\doarXiv}[1]{\href{https://arxiv.org/abs/#1}{\nolinkurl{https://arxiv.org/abs/#1}}}

\bibitem[{{Ando} {et~al.}(2019){Ando}, {Nishizawa}, {Hasegawa}, {Shimizu}, \&
  {Nagamine}}]{2019MNRAS.484.5389A}
{Ando}, R., {Nishizawa}, A.~J., {Hasegawa}, K., {Shimizu}, I., \& {Nagamine},
  K. 2019, \mnras, 484, 5389

\bibitem[{{Bagla} {et~al.}(2010){Bagla}, {Khandai}, \&
  {Datta}}]{2010MNRAS.407..567B}
{Bagla}, J.~S., {Khandai}, N., \& {Datta}, K.~K. 2010, \mnras, 407, 567

\bibitem[{Bandura {et~al.}(2014)Bandura, Addison, Amiri, Bond, Campbell-Wilson,
  Connor, Cliche, Davis, Deng, Denman, {et~al.}}]{bandura2014canadian}
Bandura, K., Addison, G.~E., Amiri, M., {et~al.} 2014, in Ground-based and
  Airborne Telescopes V, Vol. 9145, International Society for Optics and
  Photonics, 914522

\bibitem[{Battye(2013)}]{battye2013ra}
Battye, R. 2013, \mnras, 434, 1239

\bibitem[{Cai {et~al.}(2016)Cai, Fan, Peirani, Bian, Frye, McGreer, Prochaska,
  Lau, Tejos, Ho, {et~al.}}]{cai2016mapping}
Cai, Z., Fan, X., Peirani, S., {et~al.} 2016, \apj, 833, 135

\bibitem[{Cai {et~al.}(2017)Cai, Fan, Bian, Zabludoff, Yang, Prochaska,
  McGreer, Zheng, Kashikawa, Wang, {et~al.}}]{cai2017mapping}
Cai, Z., Fan, X., Bian, F., {et~al.} 2017, \apj, 839, 131

\bibitem[{{Chang} {et~al.}(2010){Chang}, {Pen}, {Bandura}, \&
  {Peterson}}]{2010Natur.466..463C}
{Chang}, T.-C., {Pen}, U.-L., {Bandura}, K., \& {Peterson}, J.~B. 2010, Nature,
  466, 463

\bibitem[{{Chang} {et~al.}(2008){Chang}, {Pen}, {Peterson}, \&
  {McDonald}}]{2008PhRvL.100i1303C}
{Chang}, T.-C., {Pen}, U.-L., {Peterson}, J.~B., \& {McDonald}, P. 2008, \prl,
  100, 091303

\bibitem[{Chen(2012)}]{chen2012tianlai}
Chen, X. 2012, in International Journal of Modern Physics: Conference Series,
  Vol.~12, World Scientific, 256--263

\bibitem[{{Chen} {et~al.}(2019){Chen}, {Mo}, {Li}, {Wang}, {Yang}, {Zhou}, \&
  {Zhang}}]{2019ApJ...872..180C}
{Chen}, Y., {Mo}, H.~J., {Li}, C., {et~al.} 2019, \apj, 872, 180

\bibitem[{Cooray \& Sheth(2002)}]{cooray2002halo}
Cooray, A., \& Sheth, R. 2002, Physics Reports, 372, 1

\bibitem[{Cui {et~al.}(2017)Cui, Knebe, Yepes, Yang, Borgani, Kang, Power, \&
  Staveley-Smith}]{cui2017large}
Cui, W., Knebe, A., Yepes, G., {et~al.} 2017, \mnras, 473, 68

\bibitem[{{d'Amico} {et~al.}(2020){d'Amico}, {Gleyzes}, {Kokron}, {Markovic},
  {Senatore}, {Zhang}, {Beutler}, \& {Gil-Mar{\'\i}n}}]{2020JCAP...05..005D}
{d'Amico}, G., {Gleyzes}, J., {Kokron}, N., {et~al.} 2020, \jcap, 2020, 005

\bibitem[{{Desjacques} {et~al.}(2018){Desjacques}, {Jeong}, \&
  {Schmidt}}]{2018PhR...733....1D}
{Desjacques}, V., {Jeong}, D., \& {Schmidt}, F. 2018, \physrep, 733, 1

\bibitem[{Diemer {et~al.}(2018)Diemer, Stevens, Forbes, Marinacci, Hernquist,
  Lagos, Sternberg, Pillepich, Nelson, Popping, {et~al.}}]{diemer2018modeling}
Diemer, B., Stevens, A.~R., Forbes, J.~C., {et~al.} 2018, The Astrophysical
  Journal Supplement Series, 238, 33

\bibitem[{Dunkley {et~al.}(2009)Dunkley, Komatsu, Nolta, Spergel, Larson,
  Hinshaw, Page, Bennett, Gold, Jarosik, {et~al.}}]{dunkley2009five}
Dunkley, J., Komatsu, E., Nolta, M., {et~al.} 2009, \apjs, 180, 306

\bibitem[{Dutton {et~al.}(2011)Dutton, Bosch, Faber, Simard, Kassin, Koo,
  Bundy, Huang, Weiner, Cooper, {et~al.}}]{dutton2011evolution}
Dutton, A.~A., Bosch, F. C. v.~d., Faber, S.~M., {et~al.} 2011, \mnras, 410,
  1660

\bibitem[{Fry(1996)}]{Fry_1996}
Fry, J.~N. 1996, The Astrophysical Journal, 461

\bibitem[{{Guha Sarkar} {et~al.}(2012){Guha Sarkar}, {Mitra}, {Majumdar}, \&
  {Choudhury}}]{2012MNRAS.421.3570G}
{Guha Sarkar}, T., {Mitra}, S., {Majumdar}, S., \& {Choudhury}, T.~R. 2012,
  \mnras, 421, 3570

\bibitem[{{Guo} {et~al.}(2020){Guo}, {Jones}, {Haynes}, \&
  {Fu}}]{2020ApJ...894...92G}
{Guo}, H., {Jones}, M.~G., {Haynes}, M.~P., \& {Fu}, J. 2020, \apj, 894, 92

\bibitem[{Guo {et~al.}(2017)Guo, Li, Zheng, Mo, Jing, Zu, Lim, \&
  Xu}]{2017ApJ...846...61G}
Guo, H., Li, C., Zheng, Z., {et~al.} 2017, \apj, 846, 61

\bibitem[{{Jeong} \& {Komatsu}(2009)}]{2009ApJ...691..569J}
{Jeong}, D., \& {Komatsu}, E. 2009, \apj, 691, 569

\bibitem[{{Khandai} {et~al.}(2011){Khandai}, {Sethi}, {Di Matteo}, {Croft},
  {Springel}, {Jana}, \& {Gardner}}]{2011MNRAS.415.2580K}
{Khandai}, N., {Sethi}, S.~K., {Di Matteo}, T., {et~al.} 2011, \mnras, 415,
  2580

\bibitem[{{Klypin} \& {Prada}(2019)}]{2019MNRAS.489.1684K}
{Klypin}, A., \& {Prada}, F. 2019, \mnras, 489, 1684

\bibitem[{Kravtsov(2013)}]{kravtsov2013size}
Kravtsov, A.~V. 2013, \apjl, 764, L31

\bibitem[{Krumholz(2013)}]{krumholz2013star}
Krumholz, M.~R. 2013, \mnras, 436, 2747

\bibitem[{Krumholz {et~al.}(2008)Krumholz, McKee, \&
  Tumlinson}]{krumholz2008atomic}
Krumholz, M.~R., McKee, C.~F., \& Tumlinson, J. 2008, \apj, 689, 865

\bibitem[{Krumholz {et~al.}(2009{\natexlab{a}})Krumholz, McKee, \&
  Tumlinson}]{krumholz2009atomic}
---. 2009{\natexlab{a}}, \apj, 693, 216

\bibitem[{Krumholz {et~al.}(2009{\natexlab{b}})Krumholz, McKee, \&
  Tumlinson}]{krumholz2009star}
---. 2009{\natexlab{b}}, \apj, 699, 850

\bibitem[{{Loeb} \& {Wyithe}(2008)}]{2008PhRvL.100p1301L}
{Loeb}, A., \& {Wyithe}, J.~S.~B. 2008, \prl, 100, 161301

\bibitem[{Lu {et~al.}(2015)Lu, Mo, \& Lu}]{lu2015galaxy}
Lu, Z., Mo, H., \& Lu, Y. 2015, \mnras, 450, 606

\bibitem[{Lu {et~al.}(2014)Lu, Mo, Lu, Katz, Weinberg, van~den Bosch, \&
  Yang}]{lu2014empirical}
Lu, Z., Mo, H., Lu, Y., {et~al.} 2014, \mnras, 439, 1294

\bibitem[{Ma {et~al.}(2015)Ma, Hopkins, Faucher-Gigu{\`e}re, Zolman, Muratov,
  Kere{\v{s}}, \& Quataert}]{ma2015origin}
Ma, X., Hopkins, P.~F., Faucher-Gigu{\`e}re, C.-A., {et~al.} 2015, \mnras, 456,
  2140

\bibitem[{{Modi} {et~al.}(2019){Modi}, {Castorina}, {Feng}, \&
  {White}}]{2019JCAP...09..024M}
{Modi}, C., {Castorina}, E., {Feng}, Y., \& {White}, M. 2019, \jcap, 2019, 024

\bibitem[{Newburgh {et~al.}(2016)Newburgh, Bandura, Bucher, Chang, Chiang,
  Cliche, Dav{\'e}, Dobbs, Clarkson, Ganga, {et~al.}}]{newburgh2016hirax}
Newburgh, L., Bandura, K., Bucher, M., {et~al.} 2016, in Ground-based and
  Airborne Telescopes VI, Vol. 9906, International Society for Optics and
  Photonics, 99065X

\bibitem[{{Nishizawa} {et~al.}(2013){Nishizawa}, {Takada}, \&
  {Nishimichi}}]{2013MNRAS.433..209N}
{Nishizawa}, A.~J., {Takada}, M., \& {Nishimichi}, T. 2013, \mnras, 433, 209

\bibitem[{{Padmanabhan} {et~al.}(2016){Padmanabhan}, {Choudhury}, \&
  {Refregier}}]{2016MNRAS.458..781P}
{Padmanabhan}, H., {Choudhury}, T.~R., \& {Refregier}, A. 2016, \mnras, 458,
  781

\bibitem[{{Padmanabhan} \& {Refregier}(2017)}]{2017MNRAS.464.4008P}
{Padmanabhan}, H., \& {Refregier}, A. 2017, \mnras, 464, 4008

\bibitem[{{Padmanabhan} {et~al.}(2017){Padmanabhan}, {Refregier}, \&
  {Amara}}]{2017MNRAS.469.2323P}
{Padmanabhan}, H., {Refregier}, A., \& {Amara}, A. 2017, \mnras, 469, 2323

\bibitem[{{P{\'e}nin} {et~al.}(2018){P{\'e}nin}, {Umeh}, \&
  {Santos}}]{2018MNRAS.473.4297P}
{P{\'e}nin}, A., {Umeh}, O., \& {Santos}, M.~G. 2018, \mnras, 473, 4297

\bibitem[{{Pritchard} {et~al.}(2015){Pritchard}, {Ichiki}, {Mesinger},
  {Metcalf}, {Pourtsidou}, {Santos}, {Abdalla}, {Chang}, {Chen}, {Weller}, \&
  {Zaroubi}}]{2015aska.confE..12P}
{Pritchard}, J., {Ichiki}, K., {Mesinger}, A., {et~al.} 2015, in Advancing
  Astrophysics with the Square Kilometre Array (AASKA14), 12

\bibitem[{{Sarkar} \& {Bharadwaj}(2018)}]{2018MNRAS.476...96S}
{Sarkar}, D., \& {Bharadwaj}, S. 2018, \mnras, 476, 96

\bibitem[{{Sarkar} {et~al.}(2016){Sarkar}, {Bharadwaj}, \&
  {Anathpindika}}]{2016MNRAS.460.4310S}
{Sarkar}, D., {Bharadwaj}, S., \& {Anathpindika}, S. 2016, \mnras, 460, 4310

\bibitem[{Skibba {et~al.}(2014)Skibba, Smith, Coil, Moustakas, Aird, Blanton,
  Bray, Cool, Eisenstein, Mendez, Wong, \& Zhu}]{Skibba_2014}
Skibba, R.~A., Smith, M. S.~M., Coil, A.~L., {et~al.} 2014, The Astrophysical
  Journal, 784, 128

\bibitem[{{Spinelli} {et~al.}(2020){Spinelli}, {Zoldan}, {De Lucia}, {Xie}, \&
  {Viel}}]{2020MNRAS.493.5434S}
{Spinelli}, M., {Zoldan}, A., {De Lucia}, G., {Xie}, L., \& {Viel}, M. 2020,
  \mnras, 493, 5434

\bibitem[{Springel(2005)}]{2005MNRAS.364.1105S}
Springel, V. 2005, \mnras, 364, 1105

\bibitem[{Springel {et~al.}(2001)Springel, White, Tormen, \&
  Kauffmann}]{springel2001}
Springel, V., White, S.~D., Tormen, G., \& Kauffmann, G. 2001, \mnras, 328, 726

\bibitem[{Tinker {et~al.}(2008)Tinker, Kravtsov, Klypin, Abazajian, Warren,
  Yepes, Gottl{\"o}ber, \& Holz}]{tinker2008toward}
Tinker, J., Kravtsov, A.~V., Klypin, A., {et~al.} 2008, The Astrophysical
  Journal, 688, 709

\bibitem[{Tinker {et~al.}(2010)Tinker, Robertson, Kravtsov, Klypin, Warren,
  Yepes, \& Gottl{\"o}ber}]{tinker2010large}
Tinker, J.~L., Robertson, B.~E., Kravtsov, A.~V., {et~al.} 2010, The
  Astrophysical Journal, 724, 878

\bibitem[{{Umeh} {et~al.}(2016){Umeh}, {Maartens}, \&
  {Santos}}]{2016JCAP...03..061U}
{Umeh}, O., {Maartens}, R., \& {Santos}, M. 2016, \jcap, 2016, 061

\bibitem[{{Villaescusa-Navarro} {et~al.}(2018){Villaescusa-Navarro}, {Genel},
  {Castorina}, {Obuljen}, {Spergel}, {Hernquist}, {Nelson}, {Carucci},
  {Pillepich}, {Marinacci}, {Diemer}, {Vogelsberger}, {Weinberger}, \&
  {Pakmor}}]{2018ApJ...866..135V}
{Villaescusa-Navarro}, F., {Genel}, S., {Castorina}, E., {et~al.} 2018, \apj,
  866, 135

\bibitem[{Wang {et~al.}(2016)Wang, Mo, Yang, Zhang, Shi, Jing, Liu, Li, Kang,
  \& Gao}]{wang2016elucid}
Wang, H., Mo, H., Yang, X., {et~al.} 2016, \apj, 831, 164

\bibitem[{{Wyithe} \& {Loeb}(2009)}]{2009MNRAS.397.1926W}
{Wyithe}, J.~S.~B., \& {Loeb}, A. 2009, \mnras, 397, 1926

\end{thebibliography}

\end{document}